\documentclass[aps,prl,amsfonts,amsmath,amssymb,reprint,twocolumn,superscriptaddress,showpacs,a4paper]{revtex4-1}

\usepackage[squaren]{SIunits}
\usepackage[usenames,dvipsnames]{color}
\usepackage[pdftex]{graphicx}
\usepackage{bm}
\usepackage{hyperref}

\usepackage{color}
\usepackage{bbm}

\newcommand{\var}[1]{\mathrm{var}(#1)}

\newcommand{\ave}[1]{\ensuremath{\langle#1\rangle}}

\newcommand{\NA}{N_{\rm A}}
\newcommand{\NL}{N_{\rm L}}

\newcommand{\Sx}{\hat{S}_{\rm x}}
\newcommand{\Sy}{\hat{S}_{\rm y}}
\newcommand{\Sz}{\hat{S}_{\rm z}}

\newcommand{\bS}{\hat{\bf S}}

\newcommand{\bN}{\hat{\bf N}}

\newcommand{\supin}{^{{\rm (in)}}}
\newcommand{\supout}{^{{\rm (out)}}}

\definecolor{mygreen}{rgb}{0,0.5,0} 
\definecolor{myblue}{rgb}{0,0,0.75} 
\definecolor{myyellow}{rgb}{0.87,0.8,0.47} 
\definecolor{mymagenta}{cmyk}{0,1,0,0.12} 

\newcommand{\gtext}[1]{{\color{mygreen}#1}}
\renewcommand{\gtext}[1]{{\color{mygreen}}}  

\newcommand{\bF}{\hat{\bf F}}
\newcommand{\Fx}{\hat{F}_{\rm x}}
\newcommand{\Fy}{\hat{F}_{\rm y}}
\newcommand{\Fz}{\hat{F}_{\rm z}}
\newcommand{\bff}{\hat{\bf f}}

\newcommand{\kA}{\kappa_1}

\newcommand{\supzero}{^{({0})}}
\newcommand{\supone}{^{({1})}}
\newcommand{\suptwo}{^{({2})}}
\newcommand{\supthree}{^{({3})}}
\newcommand{\supfour}{^{({4})}}
\newcommand{\supscat}{^{({\rm S})}}
\newcommand{\supdephase}{^{(\theta)}}
\newcommand{\supdephasep}{^{(\bar{\theta})}}
\newcommand{\supFB}{^{({\rm FB})}}

\begin{document}

\title{Feedback-cooling of an atomic spin ensemble }

\newcommand{\ICFOAddress}{ICFO -- Institut de Ciencies Fotoniques, Av. Carl Friedrich Gauss, 3, 08860 Castelldefels, Barcelona, Spain}
\newcommand{\ParisciteAddress}{Univ. Paris Diderot, Sorbonne Paris Cit\'{e}, Laboratoire Mat\'{e}iaux et Ph\'{e}nomenes Quantiques, UMR 7162, B\^{a}t. Condorcet, 75205 Paris Cedex 13, France}
\newcommand{\ICREAAddress}{ICREA -- Instituci\'{o} Catalana de Re{c}erca i Estudis Avan\c{c}ats, 08015 Barcelona, Spain}
\newcommand{\CavendishAddress}{Cavendish Laboratory, University of Cambridge, JJ Thomson Avenue, Cambridge CB3 0HE, UK}
\newcommand{\OptosAddress}{Optos, Carnegie Campus,
Dunfermline, KY11 8GR, Scotland, UK}

\author{N. Behbood}
\affiliation{\ICFOAddress}

\author{M. Napolitano}
\affiliation{\ICFOAddress}

\author{G. Colangelo}
\affiliation{\ICFOAddress}

\author{F. Martin Ciurana}
\affiliation{\ICFOAddress}

\author{R. J. Sewell}
\affiliation{\ICFOAddress}

\renewcommand{\paragraph}[1]{~\\ \noindent {\it #1} ---}
\renewcommand{\paragraph}[1]{\noindent {\it #1} ---}

\author{M.W.~Mitchell}
\affiliation{\ICFOAddress}
\affiliation{\ICREAAddress}

\date{\today}

\begin{abstract}
We describe a measurement-and-feedback technique to deterministically prepare low-entropy states of atomic spin ensembles.  Using quantum non-demolition measurement and incoherent optical feedback, we drive arbitrary states in the spin-orientation space toward the origin of the spin space.  We observe 12 dB of spin noise reduction, or  a factor of 63 reduction in phase-space volume.  We find optimal feedback conditions and show that multi-stage feedback is advantageous.  An input-output calculation of quantum noise incorporating realistic quantum noise sources and experimental limitations agrees well with the observations.  The method may have application to generation of exotic phases of ultracold gases, for example macroscopic singlet states and valence-bond solids.   \end{abstract}


\maketitle

\paragraph{Introduction}  
Cooling of many-body systems can produce new phases of matter and new collective behaviors, e.g. superfluids~\cite{BurtonN1935,KapitzaN1938} and the fractional quantum Hall effect~\cite{TsuiPRL1982}.  Traditional cooling couples the system to a low-temperature reservoir, allowing energy to leave the system and thus reducing its entropy (provided the temperature is positive~\cite{RamseyPR1956, BraunS2013}).  The search for new many-body phenomena is now actively pursued in synthetic many-body systems composed of cold atoms, with either internal or external degrees of freedom~\cite{LewensteinBOOK2012}.  Exotic phases such as valence-bond-solids~\cite{AndersonS1987} can in principle be produced by extreme cooling in tailored potentials, and valence-bond resonance has been observed on a small scale~\cite{NascimbenePRL2012}.  Scaling to larger systems places intense demands on cooling, however. Evaporative cooling~\cite{DavisPRL1995} followed by demagnetization cooling~\cite{MedleyPRL2011} has reached a record 350 pK temperature, but estimates of critical temperatures are lower still~\cite{CapogrossoPRA2010}.  Feedback cooling  An alternative proposal for producing large-scale many-body correlations employs quantum non-demolition (QND) collective spin measurements~\cite{KoschorreckPRL2010a,KoschorreckPRL2010b,SewellPRL2012} to detect  spatially-resolved spin correlations~\cite{EckertNP2008}, and sets them to desired values by feedback to the collective spin~\cite{HauckePRA2013}.  A proposal to produce macroscopic singlet states~\cite{TothJP2010,UrizarARX2012} employs the same strategy without spatial resolution.  

Here we experimentally demonstrate spin cooling by QND measurement plus feedback, using an ensemble of $^{87}$Rb held in an optical dipole trap and QND measurements~\cite{KoschorreckPRL2010a, KoschorreckPRL2010b, SewellARX2013} by off-resonance Faraday rotation probing.  The technique is similar to feedback cooling in trapped electrons \cite{DUrsoPRL2003}, nano-mechanical resonators \cite{HopkinsPRB2003, PoggioPRL2007}, single ions \cite{BushevPRL2006}, and single atoms \cite{KochPRL2010}.  In an adaptation of \cite{TothJP2010}, we start from a high-entropy state, i.e., occupying a large volume of collective spin $\bF$ phase space, and drive toward a low-entropy state, specifically toward the singlet state $\ave{\bF\cdot\bF}=0$.  We analyze this quantum control problem~\cite{MabuchiIJRNC2005} using input-output relations and ensemble-based noise models~\cite{KoschorreckJPBAMOP2009, ColangeloARX2013}, to identify an optimized two-round feedback protocol.  We demonstrate a spin noise reduction by 12 \deci\bel, or a reduction in phase-space volume by a factor of 63.

\newcommand{\supi}{^{(i)}}
\newcommand{\gauss}{G}
\newcommand{\gnaive}{g_{\mbox{\rm na\"{i}ve}}}
\renewcommand{\gnaive}{G_{0}}

\begin{figure}[t]
	\includegraphics[width=\columnwidth]{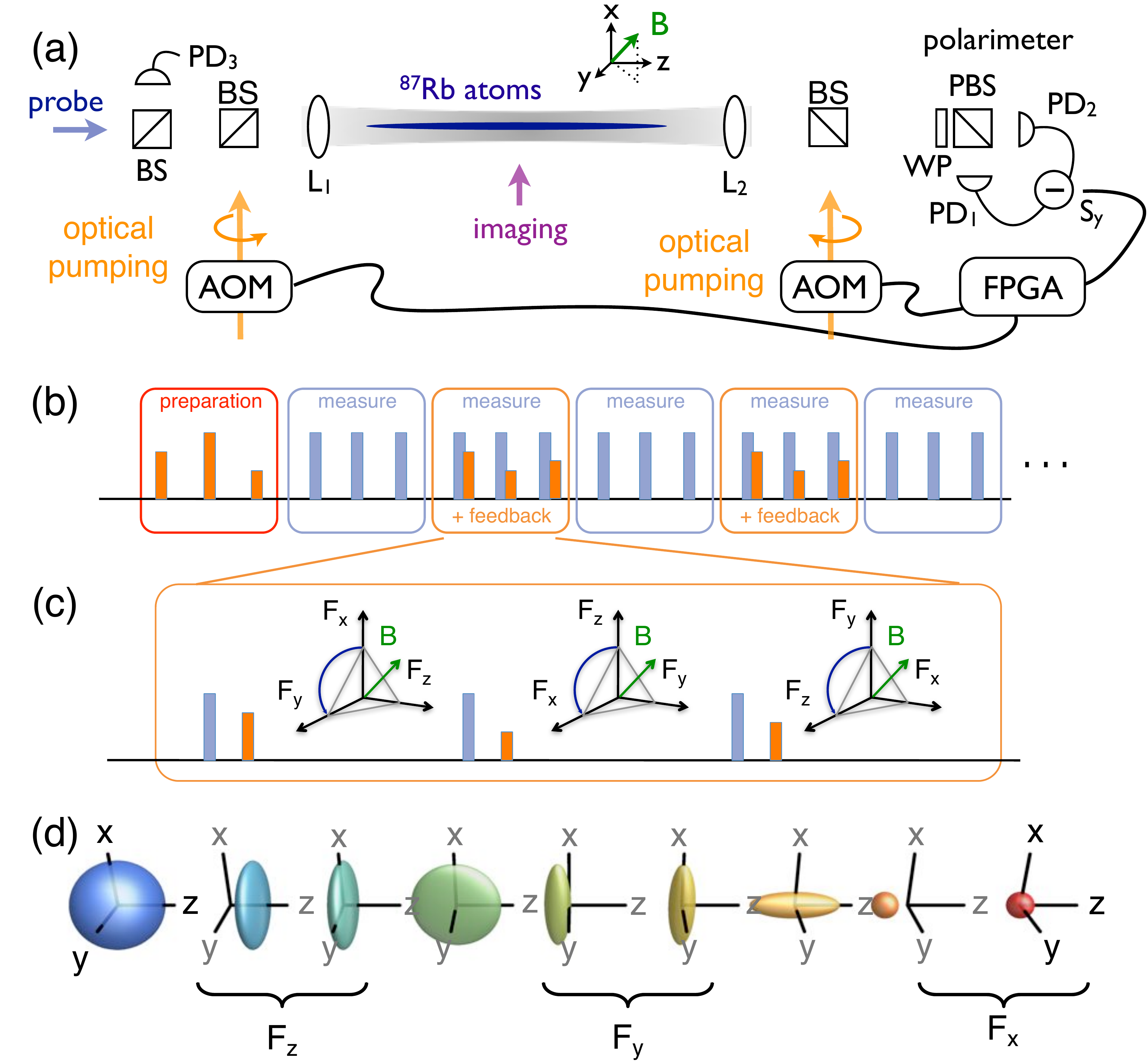}	
	\caption{Experimental schematic, pulse sequence, and control diagram for spin cooling by QND measurement + feedback. {\bf (a)} Experimental geometry.  Near-resonant probe pulses pass through a cold cloud of $^{87}$Rb atoms and experience a Faraday rotation by an angle proportional to the on-axis collective spin $\Fz$.  The pulses are initially polarized with maximal Stokes operator $\Sx$.   Rotation toward $\Sy$ is detected by a balanced polarimeter consisting of a wave-plate (WP), polarizing beam-splitter (PBS), and photodiodes (PD$_{1,2}$).  A field-programmable gate array (FPGA)-based controller interprets the polarimeter signal and Reference (PD$_3$) and produces optical feedback pulses via acousto-optic modulators (AOM)s.  $\bF$ precesses about a magnetic field (${\bf B}$) along the direction [1,1,1] making all components accessible to measurement and feedback through stroboscopic probing. {\bf (b-c)} Pulse sequence: { A first QND measurement measures the $F_z$ angular momentum component and the FPGA calculates the Faraday rotation angle in $\approx$11\unit{}{\micro\second}.  The FPGA applies a control pulse, proportional to the Faraday rotation angle, to an AOM to generate optical-pumping feedback.  At the appropriate times in the Faraday rotation cycle, the same process is applied to also to $F_y$ and $F_x$. } {\bf (d)} evolution of the state in $\bF$ phase space as successive measurement, feedback and precession steps transform the state.
	\label{fig:SetupAndControlFlow}}
\end{figure}

\paragraph{System}
The experiment is shown schematically in Figure \ref{fig:SetupAndControlFlow}(a).  Our atomic spin ensemble consists of $\NA \approx 10^6$ rubidium-87 atoms in the $f=1$ ground hyperfine level, held in an optical dipole trap elongated in the $z$ direction. \gtext{, in the presence of a magnetic field of $B \approx 5.9$ \milli\gauss~along the direction $[1,1,1]$.}  Interactions among the atoms due to collisions and magnetic dipolar couplings are negligible at our density of $\sim 10^{11}$\centi\meter$^{-3}$.  We define the collective spin operator ${\bF}\equiv\sum_{i}{\bff}^{(i)}$, where ${\bff}\supi$ is the spin of the $i$'th atom.  The collective spin obeys commutation relations $[\Fx,\Fy]=i \Fz$ (we take $\hbar=1$ throughout). \gtext{ and thus spin uncertainty relations, e.g. $\Delta F_x \Delta F_y \ge |\langle F_z \rangle|/2$. 
In the presence of the magnetic field, the ground state is a pure product state with all atoms oriented along the direction $[1,1,1]$.}  

\newcommand{\supN}{^{{\rm (n)}}}
\newcommand{\supinN}{^{{\rm (in,n)}}}
\newcommand{\supoutN}{^{{\rm (out,n)}}}
\newcommand{\subRO}{_{\rm RO}}
\newcommand{\subFB}{_{\rm FB}}
\newcommand{\supPM}{^{{\rm \pm}}}
\newcommand{\kFB}{k\subFB\supPM}

Probe pulses are described by the Stokes operator ${\bS}$ defined as  $ \hat{S}_i \equiv \tfrac{1}{2} (\hat{a}_+^\dagger,\hat{a}_-^\dagger) \sigma_i (\hat{a}_+,\hat{a}_-)^T $, where the $ \sigma_i $ are the Pauli matrices and $ \hat{a}_\pm $ are annihilation operators for $\sigma_\pm$ polarization.  As with $\bF$, the components of $\bS$ obey $[\Sx,\Sy]=i \Sz$ and cyclic permutations.  The input pulses are fully $\Sx$-polarized, i.e. with $\ave{\Sx} = \NL/2$,  $\ave{\Sy}=\ave{\Sz}=0$ and $\Delta^2 S_i = \NL/2$, $i\in\{x,y,z\}$ where $\NL$ is the number of photons in the pulse.  
While passing through the ensemble, the probe pulses experience the interaction hamiltonian
$\hat{H}_{\rm eff} =  {\kA}{\tau^{-1}}\Sz\Fz$, where $\kA$ is a coupling coefficient for vector light shifts ~\cite{Echaniz2008,ColangeloARX2013}.  This rotates the pulse by an angle $\phi=\kA \Fz \ll 1$, so that a measurement of $\Sy\supout/\Sx\supin$ indicates $\Fz$ with a shot-noise-limited sensitivity of $\Delta \Fz = \Delta \Sy /\kA$.  Tensor light shifts are negligible in this work.

\newcommand{\totvar}{\Delta^2 \bF}

\paragraph{Control strategy}
Our aim is to reduce the state's phase space volume $\Delta^2 \bF \equiv \langle |{\bF}|^2 \rangle - |\langle {\bF} \rangle|^2$ using measurement and feedback to sequentially set  $\Fz$, $\Fy$ and $\Fx$ to desired values.  This is possible using QND measurements and non-destructive feedback, which we implement with weak optical pumping.  
The spin uncertainty relations, $\Delta F_i \Delta F_j \ge |\langle F_k \rangle|/2$ even allow $\Delta^2 \bF$ to approach zero for the macroscopic singlet state  \cite{TothJP2010}.  Faraday rotation gives high-sensitivity measurement of $\Fz$.  To access $\Fx$ and $\Fy$, we apply a static magnetic field of $B \approx \unit{14}{\milli\gauss}$ along the $[1,1,1]$ axis (Larmor period $T_L\simeq\unit{120}{\micro\second}$) to induce $\Fz\rightarrow\Fy\rightarrow\Fx$ precession, and probe at $T_L/3$ intervals.  
The optical pumping performs a controlled displacement of the spin state (a rotation would leave $ |{\bF}|$ unchanged) toward a desired value.  
For increased accuracy, we repeat the three-axis measurement and feedback; the deleterious effects of measurement back-action, optical pumping noise, and feedback errors diminish when approaching ${\bF}=\bf{0}$.  The experimental sequence is illustrated in Fig.~\ref{fig:SetupAndControlFlow}(b). 

\newcommand{\outcome}{{\cal F}}

\paragraph{QND Measurement}
We measure the collective spin component $F_z$ by paramagnetic Faraday rotation probing with \unit{1}{\micro\second} long pulses of linearly polarized light with on average $\NL=5.4\times10^7$ photons per pulse at a detuning of \unit{700}{\mega\hertz} to the red of the ${\it f}=1 \rightarrow {\it f}'=0$ transition.  Measurements are made at $T_L/3\simeq\unit{40}{\micro\second}$ intervals, to access sequentially $\Fz$, $\Fy$ and $\Fx$.  A balanced polarimeter detects $\Sy\supout$ while a reference detector before the atoms detects $\Sx\supin$.  Both signals are collected by a real--time FPGA--based controller, which computes the measurement result $\outcome\equiv\Sy\supout/(\kA\Sx\supin)$ and generates timing signals to control the optical pumping feedback.  

\newcommand{\tlat}{t_{\rm lat}}

\paragraph{Optical pumping and feedback}
The optical pumping is performed in a nearly-linear regime, i.e. with few photons, such that only a small fraction of the atoms change state.   We use circularly polarized light $30$ MHz red detuned from  the ${\it f}=1 \rightarrow {\it f}'=0$ transition on the D$_2$ line with an intensity $ \sim 7$ \rm{W/m$^2$}, propagating along the trap axis and chopped into $\sim$\unit{\micro\second} pulses by acousto-optic modulators (AOMs). Two beams in opposite directions, allow rapid switching between the two circular polarizations.  As with the QND measurement, Larmor precession allows feedback to $\Fz$, $\Fx$ and $\Fy$ by $\Fz$ pumping at different points in the cycle.  In the feedback step the AOMs are gated by the FPGA after a latency of  $\tlat = \unit{11}{\micro\second}$ for computation.  The FPGA determines the polarization and pulse duration $t\subFB \propto \outcome$, which in turn determines the displacement of $\bF$.  Typical feedback pulses are $1$--$2$\unit{}{\micro\second}, i.e., much shorter than the Larmor precession period, and much longer than the \unit{\sim100}{\nano\second} rise time of the AOMs.  An independent AOM amplitude control determines the overall gain of the feedback.  

\begin{figure}[t]
	\includegraphics[width=\columnwidth]{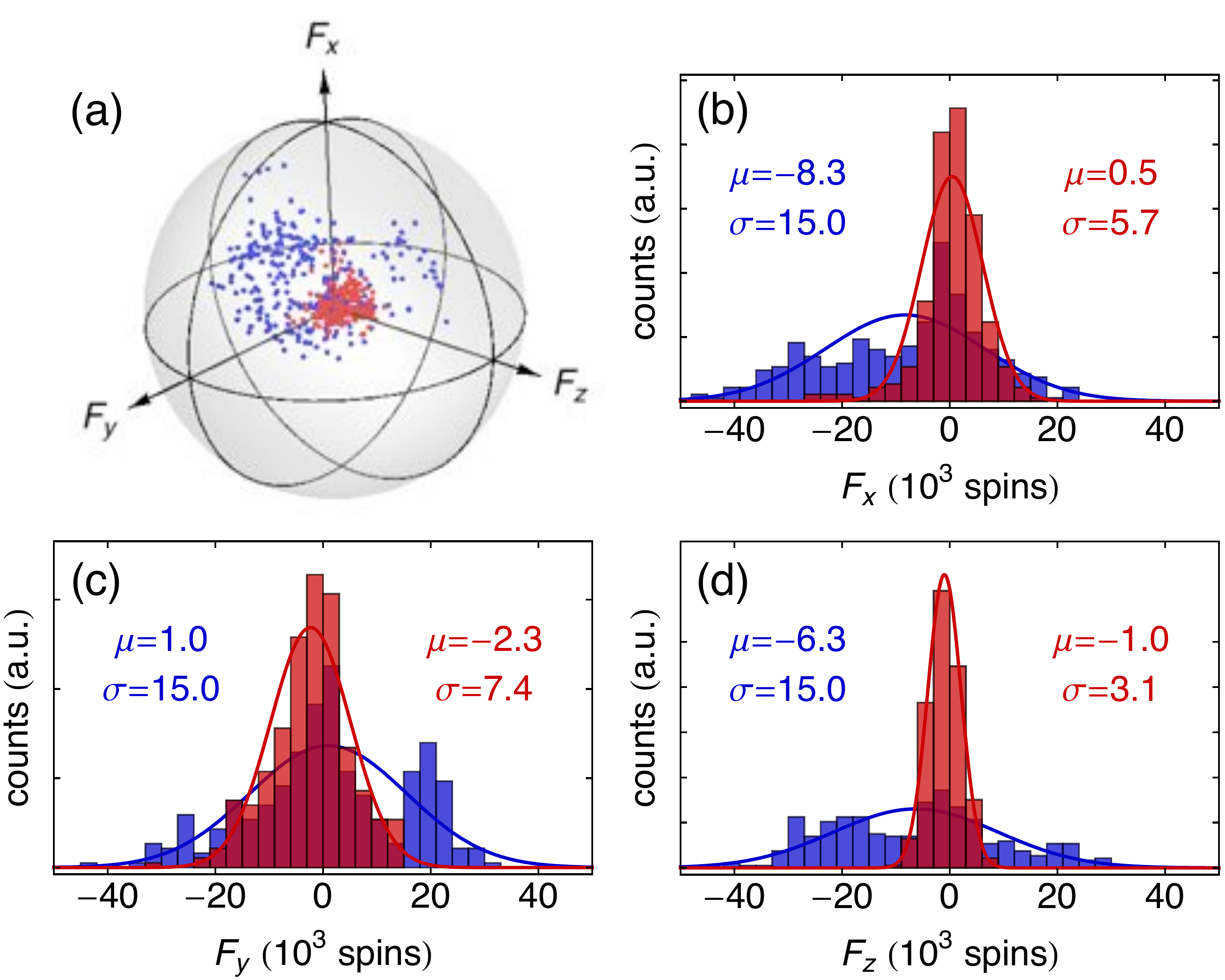}
	\caption{
	Input and output spin distributions.
	(a) Measured input spin distribution (blue data) following the initialization procedure described in the main text, and measured output spin distribution (red data) following feedback with the optimum feedback gain setting.
	The gray sphere has a radius of $6\times10^4$ spins.
	(b)--(d) Histograms of the measurements of each of the three spin components before (blue) and after (red) feedback.
	See text for details.
	\label{fig:spinDistributions}}
\end{figure}

\paragraph{Initialization procedure}
We first generate a fully mixed $f=1$ state as described in \cite{KoschorreckPRL2010a}, then optically pump $\Fz$, $\Fy$ and $\Fx$ with  \unit{5}{\micro\second} pulses.  The mixed state has zero mean and small variance $\var{F_i}=\tfrac{1}{3}f(f+1)\NA$, and serves as a fiducial point.    The amplitude and direction of the pulses are randomly chosen. The amplitude $A$ and polarization sign $s$ of the pulses are randomly chosen so that $s A $ is zero-mean normally distributed.  This generates a statistically-reproducible distribution of initial states with initial spin covariance matrix $\Gamma_F \equiv \tfrac{1}{2}\ave{F_iF_j+F_jF_i}-\ave{F_i}\ave{F_j}$ of
\begin{equation}
\Gamma_F = \left(
\begin{array}{ccc}
 2.70 & -0.03 & -1.20 \\
 -0.03 & 2.30 & -0.65 \\
 -1.20 & -0.65 & 2.20 \\
\end{array}
\right)
\times 10^8\,{\rm spins}^2,
\end{equation}
i.e., with noises $\NA^{1/2}\ll \Delta F_i \ll \NA \simeq 10^6$. 

\begin{figure}[t]
	\includegraphics[width=\columnwidth]{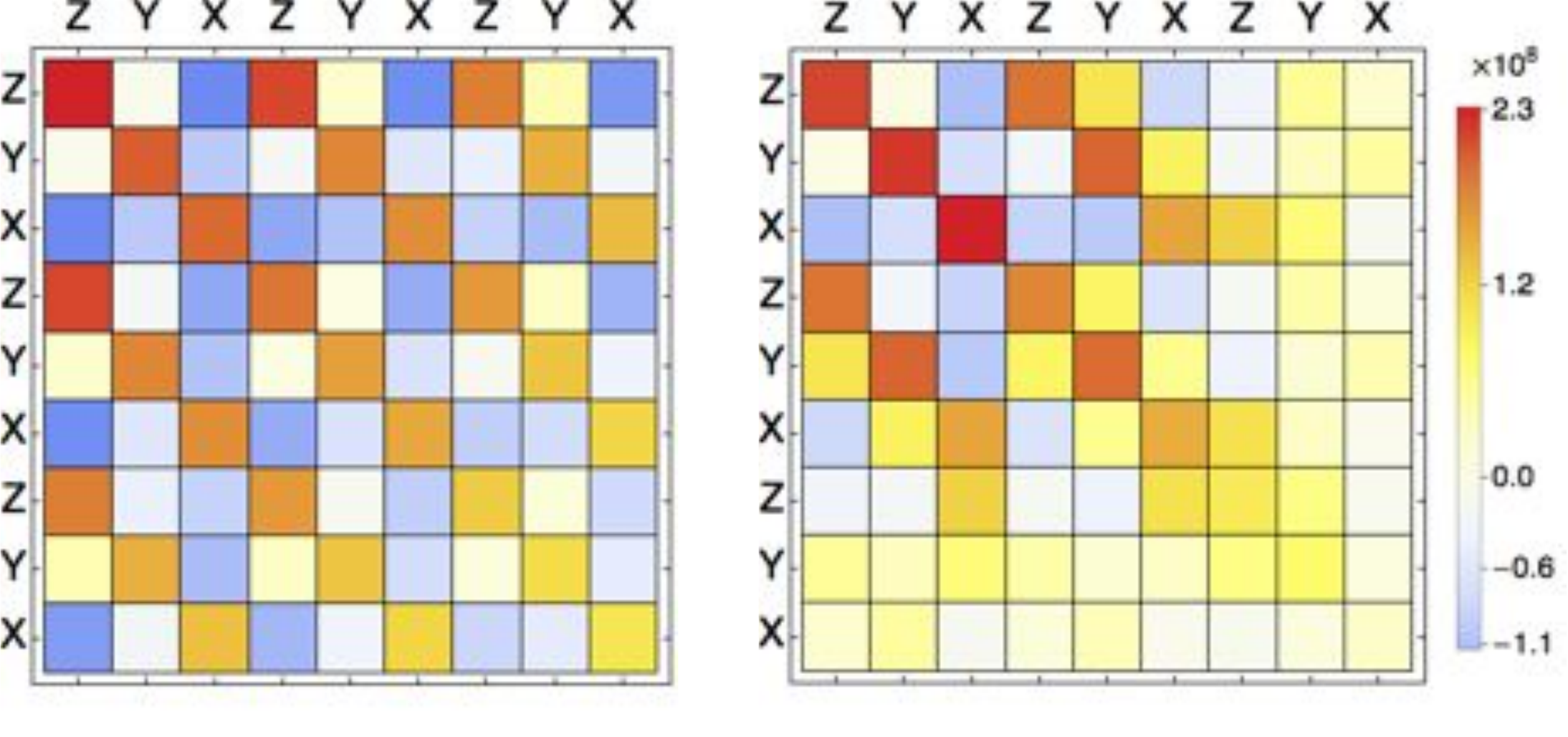}
	\caption{
	Correlation between consecutive three--component collective spin measurements.
	(a) Covariance matrix for 9 consecutive stroboscopic measurements with no feedback showing strong correlations between all three measurements of each spin component $F_i$ (red and orange squares).
	(a) Covariance matrix for 9 consecutive stroboscopic measurements with feedback after measurements 4--6 with the optimal gain setting.
	The first two measurements of each spin component $F_i$ remain strongly correlated, but the correlation is removed by the feedback and the third set of measurements is not correlated with the first two.
	Also apparent is the noise reduction after feedback.
	\label{fig:covarianceMatrices}}
\end{figure}

\paragraph{Control and characterization sequence}
For a given normalized gain $g \equiv G/\gnaive$, where $G$ is the feedback gain and $\gnaive$ is the na\"ive gain, i.e., optimal gain for the noiseless case,  we characterize the cooling process with the sequence shown in Fig. \ref{fig:SetupAndControlFlow}(b): initial state preparation, measurement without feedback, measurement with feedback, and measurement without feedback.  We then remove the atoms from the trap and repeat the same sequence to record the measurement read--out noise. The entire cycle is run 300 times to collect statistics. 

In Fig.~\ref{fig:spinDistributions}(a) we plot the input spin distribution (blue) following our initialization procedure, and the output spin distribution (red) after feedback with the optimum feedback gain setting.
The input state is distributed around the origin, with a mean deviation of $2.4\times10^4$ spins, and a total variation of $\totvar=6.7\times10^8\,{\rm spins}^2$.
Histograms of the measurements are shown in Fig.~\ref{fig:spinDistributions}(b)--(d).
After feedback (red data) the total variation of the spin distribution is $\Delta^2\bF=9.7\times10^7\,{\rm spins}^2$, an \unit{8}{dB} reduction in a single feedback step.
The dispersion of all three spin components is reduced by a factor of 3--5, and the average of each spin component remains centered within one standard deviation of the origin.

\paragraph{Correlations analysis}
Covariance matrices describing all nine measurements, for gains $g=-0.75$ (optimal case) and $g=0$ (null case) are shown in Fig. \ref{fig:covarianceMatrices}. Three features are noteworthy:  1) Both null and optimal cases show strong correlations between the first and second measurement groups, confirming the non-destructive nature of the Faraday rotation measurement.  2) the correlations of one component, e.g. $\Fy$, persist even after feedback to another component, e.g. $\Fz$, indicating the non-destructive nature of the optical feedback.  3) While the control case shows some reduction of total variance (due to spin relaxation), the feedback control is far more effective.

\begin{figure}[t]
	\includegraphics[width=\columnwidth]{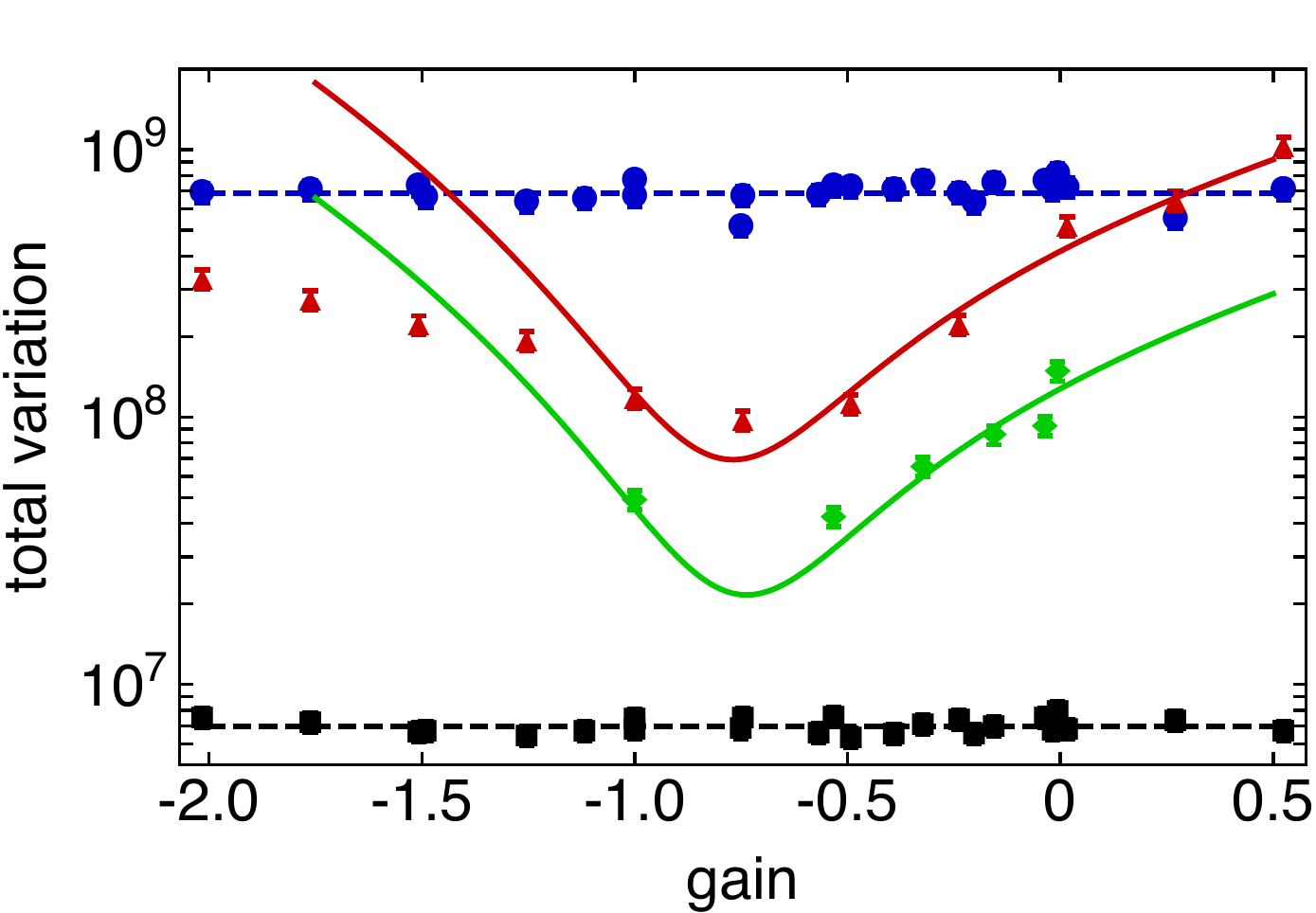}
	\caption{
	Spin cooling via optical feedback.
	Measured total variation $\Delta^2\bF$ following one (red triangles) and two (green diamonds) feedback steps.
	We compare this to the theory described in the text (red and green solid curves) fit to the $g\ge -1.0$ data as described in the text.  
	Also shown is the noise of the input spin state following the initialization procedure described in the text (blue circles), with an average total variation $\Delta^2\bF = 6.7\times10^8\,{\rm spins}^2$ (blue dashed line), and the measurement read--out noise (black squares), with an average total variation $\Delta^2\bF = 7.0\times10^6\,{\rm spins}^2$ (black dashed line).
	Error bars represent $\pm1\sigma$ statistical errors.
	\label{fig:spinCooling}}
\end{figure}

\newcommand{\cN}{{\cal N}}
\newcommand{\fzunit}{{\mathbbm{f}}_z}
\newcommand{\thetalat}{\theta}
\newcommand{\etadep}{\eta_{\rm D}}
\newcommand{\etascat}{\eta_{\rm S}}

\paragraph{Modeling}  We use a multi-step input-output model of the collective spin operators to describe the feedback cooling process.   During a step of length $\Delta t$, an operator $\hat{O}$ experiences  $\hat{O}^{(i+1)} = \hat{O}\supi - i \Delta t [\hat{O}\supi,\hat{H}_{\rm eff}\supi] + \cN$, where superscripts $\supi,^{(i+1)}$ indicate prior and posterior values, respectively, and $\cN$ is a noise operator.  Starting from atomic and optical inputs $\bF\supzero$,  $\bS\supzero$, respectively, a Faraday rotation measurement produces
\begin{eqnarray}
\label{eqn:}
\Sy\supone &=& \Sy\supzero + \kA \Sx\supin \Fz\supzero  \\
\bF\supone &=& (1-\eta) \bF\supzero - i \tau [\bF\supzero,\Fz] \Sz\supzero + \bN\supscat
\end{eqnarray}
with $\Sx,\Sz$ changing negligibly.  Measurement back-action on the atoms
$- i \tau [\bF\supzero,\Fz] \Sz\supzero$ is small provided $|\ave{\Fx}|,|\ave{\Fy}|\ll \NA$.   $\bN\supscat$ arises from the fraction $\eta$ of atoms that suffer spontaneous emission (see below).  During latency, precession by an angle $\thetalat= 2 \pi t_{\rm lat}/T_{\rm L}$ about $[1,1,1]$ causes coherent rotation $R_{\bf B}(\theta)$ and dephasing due to field inhomogeneities~\cite{BehboodAPL2013,ColangeloARX2013}:
\begin{eqnarray}
\label{eqn:rot}
{\bF}\suptwo &=& X(\theta) {\bF\supone} + {\bN\supdephase}, \\
X(\theta) & \equiv &  P_{\bf B} + \exp[{-\frac{\theta}{\omega_L T_2}}] R_{{\bf B}}(\theta) (\mathbbm{1} - P_{\bf B})
\end{eqnarray}
where $P_{\bf B}$ is a projector onto the $[1,1,1]$ direction and $T_2$ is the transverse relaxation time. 
Longitudinal relaxation is negligible on the time-scale of the experiment.  
Feedback modifies the collective spin as
\begin{eqnarray}
\label{eqn:feedback1}
{\bF}\supthree &=& G \fzunit\Sy\suptwo+ {\bF\suptwo} + \bN\supFB
\end{eqnarray}
where $G$ is the feedback  gain and $\fzunit$ is a unit vector in the $z$ direction. 
Precession by $\bar{\theta} = 2\pi/3 - \thetalat$  completes the 1/3 Larmor rotation, giving 
\begin{eqnarray}
\bF\supfour &=& X (\bar{\theta}) \left[ G\fzunit(\Sy\supzero + \kA \Sx \Fz\supzero) + X(\theta) {\bF\supzero} 
\right.
\nonumber \\ & & + 
\left.
{\bN\supFB}
+ {\bN\supdephase}+ {\bN\supscat} \right]+ {\bN\supdephasep}.
\label{eq:OneStepAffine}
\end{eqnarray}
for measurement+feedback for one component.  

The vector feedback procedure is the composition of three transformations as in Eq. (\ref{eq:OneStepAffine}). 
These correct sequentially for all three components of $\bF$, and introduce a total of twelve noise terms analogous to $\Sy\supzero$, $\bN\supFB$, $\bN\supdephase$ and  $\bN\supdephasep$, given in the Appendix.

\paragraph{Optimized multi-step cooling}
We define the normalized gain $g \equiv G/|\gnaive|$ where $\gnaive \equiv -1/(\kA \Sx)$ is the  na\"{i}ve gain, i.e., the optimal gain for zero noise, latency, and dephasing.  Minimizing ${\Delta^2\bF\supfour}$ requires $-1<g<0$ because of competition between the $G\fzunit\Sy\supzero$ and $G\fzunit \kA \Sx \Fz\supzero - X(\theta)\bF\supzero$ contributions in Eq. (\ref{eq:OneStepAffine}).  Moreover, the optimal $g$ increases with increasing signal-to-noise ratio ${\Delta^2\bF\supzero}/\Delta^2{\Sy\supzero}$.  This suggests a multi-round feedback strategy employing successive three-axis feedback steps, with decreasing $|g|$, to approach the limiting entropy set by $\Delta^2{\Sy\supzero}$ and $\Delta^2{\bN\supscat}$.  

We demonstrate this optimized multi-step cooling with results shown in Fig. \ref{fig:spinCooling}.  Again following the sequence of Fig. \ref{fig:SetupAndControlFlow}, we initialize to give measured total spin variance $\totvar\approx 6.7\times10^8\,{\rm spins}^2$, shown as blue circles.  In a first experiment we apply a single round of three-axis measurement+feedback, then measure the resulting state, and compute total variance (red triangles).   As expected, an optimum is observed at $g\simeq -0.75$, with variance $9.7\times10^7\,{\rm spins}^2$ or \unit{8}{dB} reduction in the spin noise.  In a second experiment we apply a first round with $g=-0.75$ followed by a second round with variable $g$, shown as green diamonds. This gives a further \unit{4}{dB} reduction, to $4.2\times10^7\,{\rm spins}^2$.
Model predictions, with $\kA=1.7\times10^{-7}$, $\NA=10^6$, $\NL=5.4\times10^7$, $T_2=\unit{1.3}{\milli\second}$ from independent measurements are are fit to the global data set to calibrate the optical pumping efficiency (effectively $g$), and the initial noise $\Delta^2\bF\supzero$.  Good agreement is observed except for $g \le -1.5$, a region in which the strong feedback is expected to invert and amplify the initial $\bF$.  

\paragraph{Conclusion} Using Faraday-rotation quantum non-demolition measurements and  feedback by optical pumping, we have reduced the spin variance of a laser-cooled $^{87}$Rb atomic ensemble.  The total spin variance  $\Delta^2 \bF$ is reduced by 12 dB,  or a reduction in phase-space volume by a factor of 63, using an optimized two-step procedure informed by a realistic quantum control theory incorporating experimental imperfections.  The procedure has potential application to on-demand generation of quantum-correlated states of ultra-cold atomic gases, for example generation of macroscopic singlet states and arbitrary quantum correlations in lattice-bound degenerate quantum gases.  

\paragraph{Appendix: noise terms}
Readout noise is $\Delta^2{\Sy\supzero} = \NL/2$, as above.  $\bN\supscat$ arises from spontaneous emission events, which randomize the spins of a fraction $\etascat \approx  2\kA^2 \NA\NL/(3 \alpha_0)$ of the atoms \cite{Madsen2004,KoschorreckJPBAMOP2009}, introducing a noise
$\Delta^2{N\supscat_i} = \Delta^2{f_i\supone} \etascat (1-\etascat) N_A + \etascat N_A f(f+1)/3$ where $\bff_i\supone$ is the mean single-atom spin vector.  For unpolarized states $\Delta^2{N\supscat_i}  \approx \NA \etascat(2-\etascat) f(f+1)/3$.  Similarly, dephasing randomizes the transverse polarizations of a fraction $\etadep \equiv 1-\exp[\theta/(2 \pi T_L)]$ giving noise $\Delta^2{N\supdephase_i}  \approx \NA \etadep(2-\etadep) f(f+1)/3$.  The optical pumping process is stochastic but uncorrelated among the atoms, leading to a multinomial distribution in the displacement $\bF\supFB\equiv {\bF}\suptwo - {\bF}\supone$ and a noise $\bN\supFB \propto {|\langle \bF\supFB} \rangle|^{1/2}$, which is $\ll {|\langle \bF\supFB} \rangle|$ except if ${|\langle \bF\supFB} \rangle| \sim 1$.
In this experiment with large $\alpha_0$ and small $\NL$,  only $\Sy\supzero$ and $\bN\supscat$ make a significant contribution.

\begin{acknowledgments}
We thank B. Dubost and G. T\'{o}th for helpful discussions.  This work was supported by the Spanish MINECO under the project MAGO (Ref. FIS2011-23520), by the European Research Council  under the project {AQUMET} and  by Fundaci\'{o} Privada CELLEX.
\end{acknowledgments}

\end{document}